\documentclass[useAMS,usenatbib,letterpaper]{mn2e}
\usepackage{graphicx}
\usepackage{txfonts}
\usepackage{listings}

\def\xi{\hbox{$X_{\rm i}$}}
\catcode`\@=11
\def\gsim{\ifmmode{\mathrel{\mathpalette\@versim>}}
    \else{$\mathrel{\mathpalette\@versim>}$}\fi}
\def\lsim{\ifmmode{\mathrel{\mathpalette\@versim<}}
    \else{$\mathrel{\mathpalette\@versim<}$}\fi}
\def\@versim#1#2{\lower 2.9truept \vbox{\baselineskip 0pt \lineskip 
    0.5truept \ialign{$\m@th#1\hfil##\hfil$\crcr#2\crcr\sim\crcr}}}
\catcode`\@=12
\def\msun{\hbox{$M_\odot$}}

\def\t9{\hbox{$t_9$}}

\def\m*{\hbox{$M_*$}}

\def\ho{\hbox{$H_\circ$}}
\def\h50{\hbox{$\ho /50$}}

\def\y1{\hbox{${\rm yr}^{-1}$}}

\newcommand{\aap}{A\&A}
\newcommand{\mnras}{MNRAS}
\newcommand{\apj}{ApJ}
\newcommand{\aj}{AJ}

\newcommand{\araa}{ARA\&A}

\title[Forming Globular Clusters]{Finding Forming Globular Clusters at High Redshifts}
\author[Alvio Renzini]
{Alvio Renzini$^{1}$\thanks{E-mail: $\,$alvio.renzini@oapd.inaf.it}
\\
\\
$^{1}${INAF-Osservatorio Astronomico di Padova, Vicolo
   dell'Osservatorio 5, I-35122 Padova, Italy}\\
}

\def\gtsima{$\; \buildrel > \over \sim \;$}
\def\ltsima{$\; \buildrel < \over \sim \;$}
\def\prosima{$\; \buildrel \propto \over \sim \;$}
\def\gsim{\lower.5ex\hbox{\gtsima}}
\def\lsim{\lower.5ex\hbox{\ltsima}}
\def\simgt{\lower.5ex\hbox{\gtsima}}
\def\simlt{\lower.5ex\hbox{\ltsima}}
\def\simpr{\lower.5ex\hbox{\prosima}}
\def\la{\lsim}
\def\h1{$h^{-1}$}
\def\eeq{\end{equation}}
\def\beq{\begin{equation}}
\def\24mu{24\,$\mu{\rm m}$}
\def\70mu{70\,$\mu{\rm m}$}
\def\8mu{8\,$\mu{\rm m}$}
\def\msun{\hbox{$M_\odot$}}

\def\y-1{\hbox{${\rm yr}^{-1}$}}
\def\mpc-3{\hbox{${\rm Mpc}^{-3}$}}

\begin{document}

\date{Accepted April 14, 2017; Received February  11,  2017 in original form}

\maketitle
                                                            
\label{firstpage}

\date{2016}

\begin{abstract}
The formation of globular clusters (GC) with their multiple stellar populations remains a puzzling, unsolved problem in astrophysics. One way  to gather critical insight consists in finding sizable numbers of GC progenitors (GCP)  while still near the peak of their star  formation phase, at a lookback time corresponding to GC 
ages ($\sim 12.5$ Gyr, or $z\simeq 5$). This opportunity is quantitatively explored, calculating how many GCPs could be detected by deep imaging in the optical, near-IR and mid-IR bands. For concreteness, for the imaging camera performances those of NIRCam on board of JWST are adopted. The number of GCPs that could be detected scales linearly with their mass, i.e., on how much more massive GCPs were compared to their GC progeny, and perspectives look promising. Besides 
providing direct evidence on GC formation, the detection of GCPs, their clustering, with or without a central galaxy already in place, would shed light on the relative timing of GC formation and galaxy growth and assembly. All this, may be the result of dedicated observations as well as a side benefit of deep imaging meant to search for the agents of cosmic reionization.
\end{abstract}





\section{Introduction}
\label{sec:introduction}
The formation of globular clusters (GC), with their puzzling multiple stellar populations, has emerged in recent years as a major unsolved problem in astrophysics.
Various scenarios have been proposed, in particular concerning the nature  of first generation (1G) stars, the donors of the material with the right chemical composition
for making second generation (2G) ones. However, while some scenarios  appear to encounter insurmountable difficulties in accounting for all observational constraints, none of them can claim to provide  a fully successful account  of the observational evidence (e.g., \citealt{renzini15} and references therein). Yet, the emergence of multiple stellar populations,
in a series of star formation events, each with its own peculiar chemical composition, appears to be a generic property of {\it all}  GCs, though with large quantitative differences from one cluster to another \citep{milone17}. Thus, the emergence  of multiple generations appears to be the inescapable aftermath of   the very formation of GCs. 

Perhaps the most intriguing aspect of the multiple populations phenomenon  is the so-called {\it mass budget} problem, in that the original mass in 1G stars should have been at least ten or more times higher than the mass of 1G stars still currently bound to individual clusters.
This strong requirement follows from the fact that 2G stars account for a major, sometimes dominant fraction of the  population of present-day GCs (see e.g., Figure 22 in \citealt{milone17}), whereas only a small fraction of the initial mass of 1G stars can be returned with the required  2G composition.

As a consequence, since some time the notion is widely entertained according to which  GC precursors (GCP) were substantially more massive entities than the GCs we see today in the local Universe, having been   either just more massive, oversized clusters, or even nucleated dwarf galaxies (e.g., \citealt{bekki06,dercole10,renzini13,trenti15,elmegreen17}). Still, this claim for very massive GCPs remains purely conjectural,  and even controversial. For example, \cite{larsen12} argue that the metal-poor component of the Fornax dwarf spheroidal galaxy is only $\sim 4-5$  times more massive than its GCs together, hence
claim that  no more donor mass than $4-5$  times the present mass of the GCs would have been available for the production of their 2G stars. Yet,  cosmological simulations suggest that Fornax  may have lost a substantial fraction of its initial mass \citep{wang16}. N-body simulations for the dynamical evolution of GCs have suggested that 
GCs may have been up to 10  times more massive at birth, with an average of $\sim$ 4.5 times  \citep{webb15}, though this is quite model dependent. Clearly, one definitive way of solving this controversy can best consist in finding (and weighing) GCPs near the peak of their formation phase.

The notion is also widely entertained that a substantial loss of stars from GCPs may have been a major contributor to the build up of the stellar halo of the Milky Way and perhaps even of its bulge \citep{martell11,martell16,kruissen15,schiavon17}. Clearly, this possibility opens a direct link between GC formation and the early stages in the formation and evolution of galaxies.
Indeed, GCs are the oldest objects in the local Universe for which accurate age dating has been performed, with current estimates clustering at $12.5\pm1$ Gyr
\citep{marin09,vandenberg13,brown14}, with such lookback time corresponding to $z\sim 5$, with a lower bound at $z\sim 3$ and an upper bound well into the 
reionization era. Thus, finding GCPs at these redshifts will not just help solving the puzzle of GC formation (as argued in \citealt{renzini15}), but may well shed new light on the very early stages of galaxy formation. 

On the other hand, a few candidate GCPs at $z>3$ have already been identified, thanks to strong gravitational lensing \citep{vanzella16,vanzella17a}, with a stellar mass $\sim 10^7\;\msun$ and half-light radius $\sim 60$ pc. Hence,  times are ripe to start thinking  at GCP searches on a more industrial scale.  The perspectives for detecting GCPs at these redshifts, while near  the peak of their star forming  phase, are quantitatively explored in this paper. The standard cosmology ($H_\circ=70$ km s$^{-1}$Mpc$^{-1}$, $\Omega_{\rm m}=0.3$, $\Omega_\Lambda =0.7$)  is adopted.

\section{How many GC precursors in a frame}
In the local Universe there are on average 1.5 GCs per cubic megaparsec, about 0.1 per cent of the stellar mass in galaxies is contained inside GCs and the average mass of GCs is $\sim 2\times 10^5\;\msun$ \citep{harris13}. As well known,  in most galaxies GCs show a bimodal color/metallicity distribution, with the dividing line being at 
[Fe/H]$\simeq -1.0$, with nearly equal numbers of metal-poor and metal-rich clusters \citep{harris15}. While the  local density in stars is $\sim 3\times 10^8\;\msun$ Mpc$^{-3}$ (for a Chabrier IMF), it was only $\sim 6\times 10^6\;\msun$ Mpc$^{-3}$ at $z=5$ \citep{madau14}, i.e., at the putative peak of the GC formation epoch.
Thus, if GCPs were $\gsim 10$ times more massive than their present progeny, then if already formed their mass density at $z=5$ (i.e., $10\times3\times 10^5\;\msun$ Mpc$^{-3}$) would have been of the order of $\sim 50$ per cent of the total mass in stars. The point to grasp, here, is that GCPs likely formed their stars {\it before} galaxies did a sizable fraction of their own ones, or at least so for the metal-poor half of the GC population. Indeed,  it has been even speculated that GCs may have been the first stellar systems to form after the Big Bang \citep{peebles68}, or that they may have contributed to the cosmic reionization (e.g., \citealt{schraerer11,katz13}.

Assuming that the bulk of GCs formed beyond $z=3$, one can easily calculate how many GCPs at $3<z<8$ will be sampled by an imaging camera with given field of view. For sake of concreteness one adopts here a field of view  (FoV) of 10 arcmin$^2$, close to the $4.4\times 2.2$ arcmin$^2$ total FoV of NIRCam on the {\it James Webb Space Telescope} (JWST), which is simultaneously  imaged by  its short- and the long-wavelength cameras. The possibility of a JWST detection of GCPs has been  previously discussed in the literature (e.g., \citealt{katz13,trenti15}), though with a different methodological approach.

Table 1 lists some  relevant quantities as a function of redshift, namely the cosmic time in Gyr and the comoving volume in Mpc$^3$ between $z=3$ and the other values of $z$  as sampled by a camera with a 10 arcmin$^{-2}$ FoV. This volume multiplied by 1.5 Mpc$^{-3}$ immediately gives the number of GCPs sampled by the camera in the mentioned light cone. This is a quite respectable number indeed, with almost 200,000 GCPs being sampled between $z=3$ and 8 in a single exposure. Of course, not all these GCPs will shine near their maximum luminosity, at the peak of their star formation activity, and only a small fraction of them will be bright enough to offer a chance of being detected.

Thus, let us assume GCPs best chance of being detected is when they are within 
$\sim 10$ Myr from  their peak star formation rate (SFR). From Table 1 we see that the time lapse between $z=8$ and 3 is $\sim 1.49$ Gyr,  hence  GCPs shine above the stated threshold  for only  $\sim 0.7$ per cent of the time. So, the number of GCPs observable above threshold at any given time is $\sim 1.5 \; {\rm Mpc}^{-3}\times 0.007$ times the volumes given in Table 1, or about 1 per cent of the numbers given in the third column. This is still a quite interesting number, with about 1300 GCPs being caught within 10 Myr from their peak SFR. 

Table 2 presents the expected number of GCPs in a more specific fashion,  where  one explores the efficiency in this respect of some of the NIRCam filters. Broad band (BB) filters sample the 1500 \AA\ rest-frame wavelength at the redshifts indicated in Column 4, along with the corresponding redshift interval allowed by the bandwidth. Column 5 gives the corresponding comoving volume where the filter is able to pass photons which were emitted with $\lambda=1500$ \AA.  As before, the actual number of GCPs caught within 10 Myr from their peak SFR  is $\sim 1$ per cent of the comoving volume listed in the Table. This assumes that the GC formation rate is a top-hat function of redshift, between $z=8$ and 3, hence no GCs formed at $z>8$ or $z<3$. Thus, one expects the short-wave camera to catch $\sim 60$ GCPs
still within 10 Myr from their peak SFR, and a similar number also at $z\sim 9$ if the GC formation was already under way at such higher redshift. Narrow band (NB) filters, will catch H$\alpha$ in a narrower redshift interval, thus sampling a much smaller volume (cf. Table 2) and the expectation is that the camera will catch only a few GCPs in their active phase.

As mentioned above, the star formation within GCPs did  largely precede that of the bulk of stars in galaxies, hence it may be worth comparing the number of GCPs to that of massive galaxy seeds that must inhabit the same comoving volume. For example, the local density of galaxies more massive than $10^{11}\;\msun$ is $\sim 10^{-3}$  Mpc$^{-3}$, hence from Table 1 one sees that the same $3<z<8$ comoving volume will contain the seeds for some $\sim 260$ such massive galaxies, each including of order of $\sim 500$ GCPs (could they be the {\it seeds} themselves?). Similarly, reading from Table 2 one sees that a short-wave camera exposure should sample the seeds of $\sim 6$ such massive galaxies, within the same volume where $\sim 60$ GCPs are expected to shpected to shine within 10 Myr from their maximum luminosity.

\begin{table}
\centering
{
\caption{Cosmic time, comoving volume and scale as a function of redshift.}
\begin{tabular}{lcccccc}
\hline
\hline
Redshift& Cosmic Time & Comoving Volume$^*$ & Scale\\   
               &  (Gyr)               &      (Mpc$^3)$                  &  (kpc/arcsec)\\
\hline
               8&  0.628           &      128,500                       & 4.79\\
               7&  0.750           &      109,300                       & 5.20\\
               6&  0.916           &        86,200                       & 5.69\\
               5&  1.155           &        60,500                       & 6.26\\
               4&  1.517           &        31,800                       & 6.94\\     
               3&  2.116           &       $\;\quad\quad$           0                       & 7.70\\                         
\hline
\hline
\end{tabular}
}
\label{tab:crosscheck}
{*Comoving volume in a lightcone of 10 arcmin$^2$ and between $z=3$ and $z$.\hfill}\\
From http://www.astro.ucla.edu/$\sim$wright/CosmoCalc.html
\end{table}

\begin{table}
\centering
{
\caption{Comoving volume sampled by various NIRCam filters and minimum detectable SFR.}
\begin{tabular}{lccccc}
\hline
\hline
Filter&  Flux$^\ddagger$&   Central $\lambda$&  $z\pm\Delta z^\dagger$   &Volume$^*$ &SFR\\   
          & (nJy)&             ($\mu$m)        &                                &            (Mpc$^3$)& ($\msun$/yr) \\
\hline
F070W&          20.9          &     0.704  &       3.69$\pm  0.094$    &  5,830        & 0.39\\
F090W&          14.3          &     0.902  &      5.01$\pm  0.107$     &  5,760        & 0.42\\
F115W&          11.8          &     1.154  &      6.69$\pm  0.114$      & 5,760        & 0.54\\
F150W&          11.2          &     1.501  &         9.00$\pm  0.159$   & 5,490        & 0.78\\
F212N&           265          &      2.121  &       2.231$\pm 0.006$   &   ------         &  0.66\\     
F323N&           240          &      3.237  &      3.924$\pm 0.006$   &   407           &  1.44\\                         
F405N&           260          &      4.052  &        5.173$\pm 0.006$  &  270           &  2.60\\
F470N&           341          &      4.708  &      6,152$\pm  0.005$     &    270       &  8.52\\
\hline
\hline
\end{tabular}
}
\label{tab:crosscheck}
$^\ddagger$Flux for which NIRCam will detect point-like sources with S/N=10 and 10,000 s integration.\quad\quad\quad\quad\quad\quad\quad\quad\quad\quad\quad\quad\quad\quad\quad\quad\quad\quad\quad\quad\quad\quad\\
$^\dagger$Redshift and redshift range such that $\lambda=1500$ \AA\ rest frame (BB) or H$\alpha$ \\ (NB) fall inside the bandwidth. \quad\quad\quad\quad\quad\quad\quad\quad\quad\quad\quad\quad\quad\quad\quad\quad\quad \\
$^*$Comoving volume in a lightcone of 10 arcmin$^2$ and within the sampled\\
redshift.
\quad\quad\quad\quad\quad\quad\quad\quad\quad\quad\quad\quad\quad\quad\quad\quad\quad\quad\quad\quad\quad\quad\quad\quad\quad\quad  \\
From http://www.stsci.edu/jwst/instruments/NIRCam/instrumentdesign/
\end{table}

\section{Observability}
In the previous section it has been shown that in a single exposure of NIRCam it will be plenty of GCPs, but only a fraction of them will be caught while shining near their peak brightness. Quantifying this crucial aspect is investigated in this section.

Physical arguments are given by \cite{elmegreen17} indicating in less than $\sim 3$ Myr the timescale for the bulk of stars to form within a GCP. This limit is further reinforced by the {\it chemical} constraint to avoid most of supernova contamination (the {\it supernova avoidance} constraint, \citealt{renzini15}) which imposes that most of the 1G stars are formed before core-collapse supernovae start exploding, indeed some 3 Myr after the beginning of star formation. Therefore, I assume here that first generation, 1G, stars formed within less than 3 Myr. In some scenario, where the 1G donors are massive stars (e.g., \citealt{elmegreen17}), also 2G stars are formed with very short delay, hence the whole stellar population is formed within this time limit. If the 1G donors are intermediate-mass stars (as in the \citealt{dercole10} scenario) then there should be a delay of $\sim 10^8$ yrs between 1G and 2G formation. In any event, one focuses here only on the formation of 1G stars, assuming that at formation they outnumber 2G stars.

So, critical for the number of GCPs that could be detected is their mass. As mentioned above, the average mass of present-day GCs is $\sim 2\times 10^5\;\msun$, with a Gaussian distribution in log mass, with a width of $\sim 0.5$ dex \citep{harris13}. So the typical SFR at 1G formation was:
\begin{equation}
{\rm SFR}\gsim f_{\rm MB} \frac{ M_{\rm GC}}{3\cdot 10^6{\rm yr} } \simeq 0.067\; f_{\rm MB}\frac{M_{\rm GC}}{2\cdot 10^5 \msun}\quad (\msun {\rm yr}^{-1}), 
\end{equation}
where $f_{\rm MB}$ is the mass budget factor, i.e., how much more massive were GCPs compared to their GC progeny. For example, with $f_{\rm MB} =10$, the SFR at formation of an average GCP was of order of $\sim 1\,\msun$ yr$^{-1}$ whereas that of the progenitors of the most massive GCs (such as e.g., $\omega$ Cen, with $\sim 3\times 10^6\;\msun$ today) was of the order of $10\;\msun$ yr$^{-1}$. The last column in Table 2 gives the SFR corresponding to the flux given in Column 1, using the \cite{kennicutt98} SFR-luminosity relations, for the UV continuum  or the H$\alpha$ luminosity, respectively for BB of NB filters (using the recent 
re-calibrations by \citealt{kennicutt12} detected SFRs would be $\sim 30-40$ per cent lower.

\subsection{Rest-frame 1500 \AA\ luminosity evolution of GCPs}

\begin{figure}
\vskip -2.1 cm
 \centering
 \includegraphics[width=0.52\textwidth, keepaspectratio]{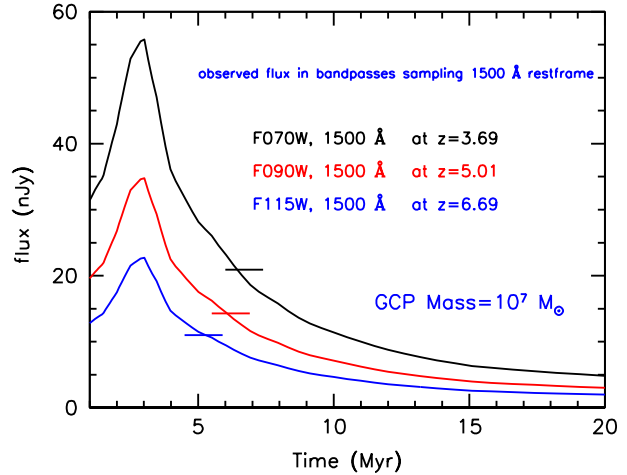}
 \vskip -2.7 cm
\caption{Rest frame time evolution of the 1500 \AA\  flux density of a $10^7\;\msun$ simple stellar population at $z=3.69$, 5.01 and 6.69 (cf. Table 2), as observed with NIRCam through the corresponding filters filters. The corresponding flux limits for a S/N = 10 detection which  are also given in Table 2.}
\label{sf}
 \end{figure}

Figure 1 shows the 1500 \AA\ rest-frame flux density evolution as a function of time for a simple stellar population of $10^7\;\msun$ (Chabrier IMF), using the SSP models of \cite{maraston05} for metallicity $Z=0.004$.  The $y$ axis gives the corresponding flux in nJy as measured on Earth through filters sampling photons emitted at 1500 \AA\ rest-frame wavelength at the indicated redshifts. The initial rise lasting $\sim 3$ Myr is due to the luminosity increase during the off-main-sequence evolution of the most massive stars, before they disappear following their supernova explosion. 

As listed in Table 2, the flux needed for a S/N =10 detection with a $\sim 3$ h integration with NIRCam ranges from $\sim 11$ nJy to $\sim 20$ nJy, respectively for the F150W and the F070W filters, as for the specific example offered on the NIRCam web page. Such a $10^7\,\msun$ GCP would be detected  above those thresholds, though for a time a bit shorter than the 10 Myr used above. This mass would correspond to a mass budget factor $f_{\rm MB}\sim 3$ for a very massive cluster like $\omega$ Cen, but $f_{\rm MB}\sim 50$ for an average GC with present mass of $2\times 10^5\;\msun$. Clearly, substantially longer integrations would be needed to detect average GCPs if the mass budget factor is, more conservatively, $\lsim 10$. For any GCP mass, the integration time needed to reach S/N=10 
can be scaled from Figure 1 as 3h times the square of $10^7/M_{\rm GCP}$, so it would be $\sim 75$h for a $M_{\rm GCP}=2\times 10^6\;\msun$ progenitor.

\subsubsection{Extinction}
These estimates do not include extinction which may affect most heavily just the peak epoch of star formation, which must occur in an extremely dense environment.
However, at least one half of  GCs are metal poor, with $-2.3\lsim{\rm [Fe/H]}\lsim -1$, hence their star formation environment should have been quite dust poor. In any event, the very first core-collapse supernovae are expected to clear  the residual gas (and dust) within a few Myr from the beginning of star formation, so quenching star formation altogether and exposing, unobscured,  the young stellar population. Thus, for metal-poor GCPs dust extinction is likely to have only a minor effect.
It is worth noting in this context that the candidate GCPs detected by \cite{vanzella17a} have indeed a very blue UV continuum, hence low extinction.

More complicated is the case for the other, metal-rich half of the GC population. Such  clusters must have formed when galaxy assembly, bulge formation and metal enrichment  have substantially proceeded (e.g., \citealt{ortolani95,shapiro10}), hence metal-rich GCPs  likely have suffered substantial extinction 
during formation, thus reducing their peak luminosity and shortening the time interval during which they could be detected.
\subsection{Other effects}

\subsubsection{Sizes}
The last column in Table 1 gives the angular scale as a function of redshift. The pixel scale is $0\prime\prime.032$ and $0\prime\prime.065$, respectively for the short- and the long-wave NIRCam cameras, which typically correspond to  $\sim 200$ pc and $\sim 400$ pc at the redshifts here considered. Objects of the size of present-day GCs (few 
ten parsecs) should appear point-like, but we don't know what  the actual size of GCPs is going to be. If they were nucleated dwarf galaxies, they may extend several hundred parsec, hence they should be marginally resolved. After all,  the local dwarfs Fornax and Sagittarius, with their share of GCs, provide evidence 
for a GC-dwarf galaxy connection still present, some 12 Gyr after formation. So, as far as we know, all GCs might have formed inside dwarfs.  However, objects similar to the lensed ones found by \cite{vanzella17a}, which have half-light radii of $\sim 60$ pc,  would be barely resolved with NIRCam.

\subsubsection{The main epoch of GC formation}
Each individual GCP will be totally dark (in UV) before experiencing its main (1G) burst of star formation, quickly reaching its peak SFR and brightness, then followed by rapid decline, possibly punctuated by  a series of secondary (2G) bursts. Then rapidly turning passive and fading well below observability. A uniform GC formation rate between $z=8$ and 3 has been assumed above, but it may well be that the GC formation rate is initially very low, then peaks at a specific redshift and then declines, with few clusters still forming at lower redshifts.  Having explored this redshift range, observations should then be able to identify 
such a {\it golden age} of GC formation, therefore establishing the relative timing of GC and galaxy formation, and providing, incidentally, a straight check for 
local GC ages as inferred from isochrones.

\subsubsection{Clustering}
Globular clusters are not uniformly distributed in the local Universe, but clusters around their {\it parent} galaxies. This must be the case also at high redshift, while they are forming, though the {\it parent} galaxy may not be in place yet. Thus, one expects candidate GCPs to be highly clustered, lying within groups of $\lsim 100$ kpc, or $\lsim 20\prime\prime$ size (see Table 1). With sufficient statistics, it should be possible to directly establish the connection between the peak of GC formation and the stage of development of the host galaxy: clusters of GCPs  as signposts of incipient massive galaxy formation.

\section{Discussion and Summary}

GCPs identified  via broad-band photometry will need a spectroscopic confirmation, and the most obvious and timely follow up will be with JWST itself, either with
the NIRCam grizm capability or with NIRSpec. In the longer run, 30-40m class telescopes with their adaptive optics capabilities will offer unparalleled resolution. However, even before spectroscopic observations, some confirmation may come for those GCP candidates detected in more than one band, with SED fits providing age, mass and  redshift constraints. Of course, more information  could be gathered for those objects detected with narrow-band filters, but they could be at most just a few.

Designing a specific observing plan is far beyond the scope of this paper. From the estimates  given here it is relatively easy to infer what kind of information could be obtained for given invested observing time.  On the other hand, deep multiband imaging over many fields, either primary or parallel, will certainly be done with JWST. This exercise has demonstrated that GCPs {\it can} be detected, with the number of detections critically depending on the mass of the progenitors, i.e., on the mass budget factor $f_{\rm MB}$. If GCPs were just marginally more massive than their present-day progeny then only the most massive ones could be detected, i.e., would have broad-band  fluxes  at formation above the thresholds given in Table 2, scaled to the actual integration time.
If this were the case, than the massive (dwarf galaxy) precursor hypothesis would not be the solution of the puzzle of globular cluster formation along with
their  multiple stellar populations. Totally different  scenarios from those so far proposed would have to be invented.

However, if observations were to find sizable numbers of GCP candidates, such as those expected if the mass budget factor is of order of a factor of $\sim 10$ or more, then a very rich field of investigation would open. Still, even in the most optimistic case just of order of $\sim 50$ GCPs could be detected above threshold in a single frame
and their images will be confused among many thousand galaxies at all possible redshifts, hence sorting them out will be a sort of {\it needle in the haystack} problem.

In summary, assuming adequate telescope time will be invested, the actual number of detected  GCPs will set constraints  and provide unique insight on:

\begin{itemize}
\item
The mass of GCPs, as the number of detections scales linearly with it, hence on the mass budget issue, along with  its consequences, such as the contribution of the
GCP dissolution to the build up of the stellar halos of galaxies.

\item
The duration of the primary (1G) star formation event, hence helping to understand the formation process itself.

\item
The size, and nature of the precursors, whether an oversized globular cluster or a nucleated dwarf galaxy.

\item
The peak epoch (golden age) of GC formation, thus setting more precise, independent  estimates for their ages

\item
The relative timing of GC formation and main galaxy assembly, from the clustering of GCPs, with or without a central galaxy already in place.
With sufficient statistics, it would be possible to estimate  the duration of the GC formation epoch within a given GC system/protogalaxy,

\end{itemize}

These are all very attractive, yet conjectural  expectations and the best way to check whether they are realized in Nature could soon be offered by deep NIRCam  imaging. While identifying the agents of reionization was a major science driver for JWST, and still is, finding sizable numbers of GCs in formation may well be a quite beneficial side effect. Actually, some very deep imaging potentially able to detect GCPs already exists, such as the {\it Hubble Ultra Deep Field} (HUDF) \citep{beckwith06}, even reaching below $\sim 5$ nJy. Some of the objects detected in the HUDF by  \cite{vanzella17b} may actually  qualify as candidate GCPs, such as objects
\# 1 and \# 7 in their Figure 5 and the two objects at $z=5.13$ in their Figure 6 (E. Vanzella, private communication). In this context, if most GCs were to form at $z\gsim 8$, then finding GCPs and the agents of reionization would become very tightly connected, both from the astrophysical and the observational points of view. 

 Before concluding, a few caveats are in order. GC ages of $12.5\pm 1$ Gyr have been used so to concentrate on the  $3<z<8$ redshift range. However, significantly younger GCs
appear also to exist (e.g., \citealt{forbes10}), hence with formation redshifts $<3$. They are not considered in this paper for two reasons: 1) the shortest wavelength filter on NIRCam will include the rest-frame 1500 \AA\ emission only down to $z\sim 3.6$, hence its capability to detect GCPs within $<10$ Myr from their peak luminosity rapidly drops towards lower redshifts, and 2) most of those definitely  younger GCs ($< 10$ Gyr) have masses well below the $2\times 10^5\,\msun$ peak, hence could be detected --from either ground or space-- only if their mass budget factor was  very large ($f_{\rm MB}\gg 10$). It is also conceivable that some massive cluster coeval with GCPs may have completely dissolved by $z=0$ (which formally would correspond to $f_{\rm MB}\sim\infty$), hence such {\it false positive}  detections may  complicate the interpretation of the actual counts of `GCPs', with  those guessed in this paper that should be taken as a lower limit along with the estimated $<\! f_{\rm MB}\!>$. In any event, the best chance of detection is obviously offered by the progenitors of GCs in the more massive half of the present day GC distribution.  In this paper the widely used first/second generation (1G/2G) nomenclature has been adopted to qualify the multiple population phenomenon, which is to say that 2G stars  formed from materials processed by 1G stars, which is in common to most scenarios so far proposed. Not everybody may agree with such nomenclature, though a viable scenario for the formation of multiple populations within a single generation does not appear to exist at the moment (e.g., \citealt{bastian16,renzini15}).

One final opportunity is worth mentioning. Complementing the direct census of GCPs at high redshifts, a cross check of the mass budget issue may  soon come from ongoing surveys such as 
ESO-Gaia \citep{randich13}, APOGEE \citep{schiavon17} or GALAH \citep{desilva15} that could detect chemically (and kinematically) tagged stars coming from the stripped GCPs of the Milky Way. If GCPs lost over $\sim 90$ per cent of their original mass, then over $\sim 20$ per cent of the Galactic halo should be made by their debris.

\section*{Acknowledgments}
I wish to thank Emanuele Daddi, Chiara Mancini,  Claudia Maraston, Lucia Pozzetti and Giulia Rodighiero for having run on my request some of their {\it scripts}, so to complete and check the two tables and the figure of this paper, and to Eros Vanzella for constructive comments and suggestions. I'm also grateful to Giampaolo Piotto and his {\it HST Treasury} Team, including Jay Anderson, Tom Brown, Santi Cassisi, Franca D'Antona, Antonino Milone and many others, for having rejuvenated my interest on globular clusters, when they turned far more complex and fascinating objects than
I  ever suspected before. I acknowledge support from a grant  INAF-PRIN 2014.

\label{lastpage}

\end{document}